\documentclass[prl,nofootinbib,floats,aps,twocolumn,preprintnumbers]{revtex4}

\usepackage[hypertex]{hyperref}
\usepackage{graphicx} 
\usepackage{xspace,amsmath}
\usepackage{enumerate}
\usepackage{tabularx}
\usepackage{bigstrut}
\usepackage[hypertex]{hyperref}

\def\be{\begin{equation}}
\def\ee{\end{equation}}
\def\ba{\begin{eqnarray}}
\def\ea{\end{eqnarray}}
\def\ge{\mathrel{\raise.3ex\hbox{$>$\kern-.75em\lower1ex\hbox{$\sim$}}}}
\def\la{\mathrel{\raise.3ex\hbox{$<$\kern-.75em\lower1ex\hbox{$\sim$}}}}

\def\simgt{\mathrel{\raise.3ex\hbox{$>$\kern-.75em\lower1ex\hbox{$\sim$}}}}
\def\simlt{\mathrel{\raise.3ex\hbox{$<$\kern-.75em\lower1ex\hbox{$\sim$}}}}

\newcommand{\nc}{\newcommand}

\nc{\gone}{\bar g_{\pi NN}^{(1)}}
\nc{\gzero}{\bar g_{\pi NN}^{(0)}}
\nc{\al}{\alpha}
\nc{\ga}{\gamma}
\nc{\de}{\delta}
\nc{\ep}{\epsilon}
\nc{\ze}{\zeta}
\nc{\et}{\eta}
\nc{\ka}{\kappa}
\nc{\rh}{\rho}
\nc{\si}{\sigma}
\nc{\ta}{\tau}
\nc{\up}{\upsilon}
\nc{\ph}{\phi}
\nc{\ch}{\chi}
\nc{\ps}{\psi}
\nc{\om}{\omega}
\nc{\Ga}{\Gamma}
\nc{\De}{\Delta}
\nc{\La}{\Lambda}
\nc{\Si}{\Sigma}
\nc{\Up}{\Upsilon}
\nc{\Ph}{\Phi}
\nc{\Ps}{\Psi}
\nc{\Om}{\Omega}
\nc{\ptl}{\partial}
\nc{\del}{\nabla}
\nc{\ov}{\overline}
\nc{\newcaption}[1]{\centerline{\parbox{15cm}{\caption{#1}}}}
\nc{\us}{U(1)$_S$}

\def\beq{\begin{equation}}
\def\eeq{\end{equation}}
\def\bmat{\begin{displaymath}}
\def\emat{\end{displaymath}}
\def\bear{\begin{eqnarray}}
\def\eear{\end{eqnarray}}
\def\ba{\begin{eqnarray}}
\def\ea{\end{eqnarray}}
\def\bery{\begin{array}}
\def\ery{\end{array}}
\def\bit{\begin{itemize}}
\def\eit{\end{itemize}}
\def\ben{\begin{enumerate}}
\def\een{\end{enumerate}}
\def\btab{\begin{tabular}}
\def\etab{\end{tabular}}
\def\btbl{\begin{table}}
\def\etbl{\end{table}}
\def\bfig{\begin{figure}[htb]}
\def\efig{\end{figure}}
\def\bpic{\begin{picture}}
\def\epic{\end{picture}}

\def\ga{\mathrel{\raise.3ex\hbox{$>$\kern-.75em\lower1ex\hbox{$\sim$}}}}
\def\la{\mathrel{\raise.3ex\hbox{$<$\kern-.75em\lower1ex\hbox{$\sim$}}}}
\def\gappeq{\mathrel{\rlap {\raise.5ex\hbox{$>$}}
{\lower.5ex\hbox{$\sim$}}}}
\def\lappeq{\mathrel{\rlap{\raise.5ex\hbox{$<$}}
{\lower.5ex\hbox{$\sim$}}}}

\def\gyr{{\rm \, G\kern-0.125em yr}}
\def\mev{{\rm \, Me\kern-0.125em V}}
\def\gev{{\rm \, Ge\kern-0.125em V}}
\def\tev{{\rm \, Te\kern-0.125em V}}

\begin{document}

\pagestyle{plain}  

\preprint{PI-PARTPHYS-192}

\title{\boldmath Dark Discrete Gauge Symmetries}

\author{Brian Batell}

\affiliation{Perimeter Institute for Theoretical Physics, Waterloo,
ON, N2J 2W9, Canada}


\begin{abstract}

We investigate scenarios in which dark matter is stabilized by an abelian $Z_N$ discrete gauge symmetry. 
Models are surveyed according to symmetries and matter content.
Multi-component dark matter arises when $N$ is not prime and $Z_N$ contains one or more subgroups.
The dark sector interacts with the visible sector through the renormalizable kinetic mixing and Higgs 
portal operators, and we highlight the basic phenomenology in these scenarios. In particular, 
multiple species of dark matter can lead to an unconventional nuclear  recoil spectrum in direct detection experiments,
while the presence of new light states in the dark sector
can dramatically affect the 
decays of the 
Higgs at the Tevatron and LHC, thus providing a 
window into the gauge origin of the stability of dark matter.

\end{abstract}

\maketitle


 
There is compelling empirical evidence for the existence of dark matter (DM) in the Universe~\cite{review}. 
A simple and attractive possibility for DM is a new elementary particle that is stable on
cosmological times scales.  
This stability strongly suggests the existence of a new symmetry in the dark sector, 
and understanding the precise nature of this symmetry is of fundamental importance in our quest for a 
theory of DM. 

While there are many possibilities for symmetries that stabilize DM, by far the most common example invoked 
in the literature is a discrete $Z_2$ symmetry, e.g. R-parity in Minimal Supersymmetric Standard Model (MSSM).
Generic parity symmetries are clearly
a useful tool in building models of DM, but they often appear ad hoc. Indeed, discrete symmetries are not an organizing 
principle of the Standard Model (SM), and in general quantum field theories suffer no theoretical inconsistency once 
discrete symmetries are abandoned. It has also been argued that global discrete symmetries will be violated
by quantum gravitational effects~\cite{discrete}. It is therefore reasonable to ask if discrete symmetries may find their 
origin as a consequence of a more fundamental principle that may ultimately be responsible for the stability of DM. 

Gauge symmetry is just such a principle that may provide the eventual rationale for discrete symmetries. 
A so-called discrete gauge symmetry emerges as a remnant of a spontaneously broken gauge symmetry \cite{discrete}. 
These symmetries are exactly conserved owing to their gauge origin, 
and thus the lightest state having a nontrivial charge under the discrete symmetry is absolutely stable. In the past, 
discrete gauge symmetries have been used in a variety of contexts, but most notably 
to explain or find an alternative to $R$-parity in the MSSM~\cite{Rparity}. 
Very few studies have been solely motivated by explaining DM stability, 
but see Ref.~\cite{walker} for a recent example based on a $Z_2$ remnant 
from a nonabelian SU(2) gauge symmetry. See Ref.~\cite{nonabelian} for earlier work on dark matter models
with nonabelian discrete gauge symmetries.

In this paper we will investigate models of hidden sector DM based on a spontaneously broken 
U(1)$_D$ gauge symmetry which preserves a discrete $Z_N$ subgroup.  While U(1)$_D$ is the minimal choice 
for the dark sector gauge group, it also affords the possibility of renormalizable interactions of 
DM with the SM via the kinetic mixing~\cite{holdom} and Higgs portal~\cite{singlet1} operators. 
New phenomena may occur for large $Z_N$ symmetries. This includes multiple 
stable DM candidates, new direct detection signatures, and novel decays of the SM Higgs boson leading to 
final states with multiple leptons, jets, and missing energy. These signatures provide a means 
to probe the gauge origin of DM stability.


{\bf Discrete Gauge Symmetry.} 
Let us begin by reviewing the prototype setup for a gauged $Z_N$ 
discrete symmetry \cite{discrete}. Consider a U(1)$_D$ gauge theory containing a dark Higgs field $\phi_{N}$ 
with integer charge $N$ and scalar matter field $\chi_{1}$ with
charge 1, where the notation is that the subscript indicates the charge of the field under U(1)$_D$. 
In addition to the usual self-conjugate terms the gauge symmetry allows the following term in the Lagrangian:
\begin{equation}
\Delta {\cal L} \propto  \phi^\dag_{N} \chi_{1}^N + ~{\rm h.c.}
\label{Dterm}
\end{equation} 
We assume the scalar potential is constructed so that the dark Higgs gets a vacuum expectation value (vev). 
This spontaneously breaks the gauge symmetry, but the Lagrangian preserves a
discrete $Z_N$ subgroup of U(1)$_D$ under which the matter fields transform as 
\begin{equation}
\chi_1 \rightarrow e^{2\pi i k/N} \chi_1, \qquad k= 0,1,2,\dots N-1.
\label{zn}
\end{equation}
In general, as long as there are no additional sources of gauge symmetry breaking and all gauge anomalies are canceled 
with appropriate matter content, the remnant discrete symmetry will be exactly conserved. The case $N=2$ corresponds 
to the usual parity so often encountered in models of DM, but in this 
construction there is nothing particularly special about the choice of $Z_2$, as it is simply a consequence of 
the relative charges of the Higgs and matter fields in the dark sector which are a priori undetermined.
Note that depending on the field content and associated gauge and Lorentz charges, 
there may be no renormalizable terms analogous to Eq.~(\ref{Dterm}) present. 
In this case there will be accidental global symmetries which also may stabilize DM, 
at least at the renormalizable level. Of course these accidental symmetries may be violated at the nonrenormalizable 
level, in contrast to the discrete gauge symmetry. 
 

{\bf Models.} We now give a survey of DM models based on an abelian gauged $Z_N$ symmetry, 
generalizing Eq. (\ref{Dterm}) to include fermions, multiple matter fields and chiral matter.  
Models are characterized by the field content and corresponding charges. The full Lagrangian 
contains kinetic terms for the U(1)$_D$ gauge boson $V_\mu$ and the dark Higgs boson $\phi_{N}$,
as well as an appropriate scalar potential such that  $\phi_{N}$ acquires a vev 
$\langle \phi_{N} \rangle = v'/\sqrt{2}$. This vev breaks the gauge symmetry leaving a massive vector and
a physical dark Higgs fluctuation $h'$. In addition there are gauge kinetic mixing and Higgs portal terms that
allow interactions with the SM; we will discuss these in detail below. Finally, there will be
model dependent terms related to the additional matter fields which comprise the DM sector.
We begin by examining models with minimal matter content, i.e. one dark Higgs and one matter field, leading to three 
possibilities at the renormalizable level: 1) a $Z_2$ scalar, 2) a $Z_2$ Dirac fermion, 
and 3) a $Z_3$ scalar. We then turn to multi-field and chiral models.


{\it $Z_2$ scalar:} The $Z_2$ scalar DM model contains a charge 2 dark Higgs field $\phi_{2}$ 
and a charge 1 scalar matter field $\chi_{1}$. The gauge symmetry permits the term (\ref{Dterm}) 
to appear in the Lagrangian:
\begin{equation}
 \Delta {\cal L}  = - \frac{\lambda}{\sqrt{2}} \phi^\dag_2 \chi_1 \chi_1 +{\rm h.c.} , \\
\label{z2S}
\end{equation}
where a possible phase in the coupling $\lambda$ may be removed by rotating $\chi_1$. 
A novel aspect of $Z_2$ models is that Eq.~(\ref{z2S}) induces a mass splitting between the real and imaginary 
components of the matter field, $ \chi_1 =  (S + iP)/\sqrt{2}$, after U(1)$_D$ is spontaneously broken.
Thus, the physical masses of the $S$ and $P$ scalars are
$m_{S,P}^2 = m_0^2 \mp \Delta m^2$
where $m_0^2$ includes universal contributions from a bare mass operator $\chi_1^\dag \chi_1$ and Higgs portal
operators $(\phi_2^\dag \phi_2) (\chi_1^\dag \chi_1)$, and $(H^\dag H) (\chi_1^\dag \chi_1)$ following gauge symmetry breaking, 
while $\Delta m^2 \equiv \lambda v'$ comes from Eq.~(\ref{z2S}).
According to Eq.~(\ref{zn}) the components of the multiplet are $Z_2$ odd, 
$S \rightarrow - S$, $P\rightarrow -P$, so that for negative $\lambda$, $S$ will be the stable DM candidate.

The splitting of the individual components $S,P$ leads to an 
off-diagonal interaction coming from the gauge kinetic term of the matter field:
\begin{equation}
 g_D V_\mu P \overleftrightarrow {\partial_\mu} S, 
\label{VPS}
\end{equation}
where $g_D$ is the gauge coupling of U(1)$_D$.
One well-known consequence of this operator is that scattering with nuclei mediated by vectors 
$V_\mu$ becomes inelastic \cite{inelastic}. 
Another important effect is that this interaction will inevitably allow the heavier $Z_2$ partner 
$P$ to decay to the DM plus light SM fermions which will have important implications for collider physics,  
as we will discuss in detail later. 


{\it $Z_2$ fermion.} Consider next the minimal theory of fermionic DM stabilized by a discrete $Z_2$ gauge symmetry. The 
model contains a charge 2 dark Higgs $\phi_{2}$, and a vectorlike pair of Weyl fermions $\psi_{1}$ and 
$\xi_{-1}$ with charges +1 and -1, respectively. Besides the kinetic terms and Dirac mass term, 
${\cal L} \supset - m_D (\psi_1 \xi_{-1} +{\rm h.c.})$, 
the gauge symmetry permits the following terms analogous to Eq.~(\ref{Dterm}):
\begin{equation}
\Delta {\cal L} = -\frac{\lambda_L}{\sqrt{2}} \phi_2^\dag \psi_1 \psi_1 
-\frac{\lambda_R}{\sqrt{2}} \phi_2 \xi_{-1} \xi_{-1} +{\rm h.c.}  \\
\label{z2F}
\end{equation}
Like the $Z_2$ scalar above, there is a splitting between the physical mass eigenstates states.
which can be written in terms of Majorana fermions $\Psi_{\pm}$, 
with masses $m_{\pm} \simeq m_D \pm (m_L +  m_R)/2$. Here we have defined $m_{L,R} \equiv \lambda_{L,R} v'$ and 
made the assumption $m_D \gg m_{L,R}$ for simplicity. There is discrete $Z_2$ gauge symmetry that remains under which both 
$\Psi_{\pm}$ are odd, ensuring the stability of the lightest particle. 
In the limit where $\lambda_L = \lambda_R$, the gauge interaction becomes
 \begin{equation}
  i  g_D V_\mu \, \overline{\Psi}_+ \gamma^\mu \Psi_- .  
\label{V12}
\end{equation}
We see that, at least structurally, the $Z_2$ fermion has many similar features to the $Z_2$ scalar discussed above,
and in particular the gauge interaction is off-diagonal. We note that both the $Z_2$ scalar and $Z_2$ fermion models 
were considered in the past as models of inelastic DM~\cite{inelastic,inelastic2}, 
and a similar model with two Dirac fermions was examined in Ref.~\cite{hidferm}.


{\it $Z_3$ scalar. } The final model with minimal field content is the the $Z_3$ scalar DM model. This
model contains a charge 3 Higgs field $\phi_{3}$ and a charge 1 scalar matter field $\chi_1$. 
The gauge symmetries allow the additional term (\ref{Dterm})
\begin{equation}
 \Delta{\cal L}  =   -\frac{\sqrt{2}}{3!}\, \lambda \, \phi^\dag_3 \chi_1 \chi_1 \chi_1 + {\rm h.c.} \\ 
\label{z3}
\end{equation}
Unlike the $Z_2$ models considered above, there is no mass splitting between the real and imaginary components, which 
clearly must be the case since $Z_3$ charged fields are necessarily complex. After symmetry breaking, the interactions
from (\ref{z3}) become
\begin{equation}
    -\frac{1}{3!} \lambda v'  \chi_1\chi_1\chi_1   -\frac{1}{3!} \lambda   h'  \chi_1\chi_1\chi_1 +{\rm h.c.} 
\label{z3int}
\end{equation}
The interaction with the dark Higgs $h'$ leads to the possibility of `semi-annihilation', $\chi_1 \chi_1 \rightarrow \chi_1 h'$,  
and may have interesting consequences for the thermal history of $\chi_1$, as emphasized recently in \cite{semiA}. Note that
the interactions in (\ref{z3int}) being cubic in $\chi_1$, do not allow for a tree-level  DM-nucleus scattering. 
Direct detection is of course still possible through other portal interactions. 
For other models of $Z_3$ DM, see Ref.~\cite{z3models} 


{\it Multi-field $Z_N$ models.} With a single matter field, as considered in the models above, 
the renormalizable possibilities for discrete gauge symmetries are limited to $Z_2$ and $Z_3$.  
Once we allow for more than one matter field in the dark sector, renormalizable $Z_N$ models are possible. 
Here we will consider several interesting features of these models, illustrating them with examples.

Generically, if $N$ is a prime number then barring accidental global symmetries there will be only one DM candidate: 
the lightest particle with nontrivial $Z_N$ charge\footnote{Additional heavier particles with $Z_N$ charge can be stable
if all possible decay modes are forbidden by kinematics.}. However, if $N$ is not prime, but rather composite 
(has divisors besides 1 and $N$), 
then the Lagrangian will preserve
additional discrete symmetries that are subgroups of $Z_N$, allowing for the possibility of multiple stable species 
and thus multiple DM candidates~\cite{LUP}. Note that multi-component DM has been considered for a variety of 
reasons in the past; see Ref.~\cite{MCDM}. To illustrate, consider the simplest case in which this occurs, namely a 
$Z_4$ symmetry, which has $Z_2$ as a subgroup. A multi-component DM model may be constructed with the following field content:
a charge 4 dark Higgs  $\phi_{4}$ and scalar matter fields $\chi_{2}$ and
$\chi_{1}$ with charges 2 and 1. The Lagrangian contains
\begin{eqnarray}
\Delta{\cal L} & =& -\lambda_1 \phi^\dag_{4} \chi_{2} \chi_{2}  
-\lambda_2 \phi^\dag_{4} \chi_{2} \chi_{1}\chi_{1}  
- \lambda_3 \chi^\dag_{2} \chi_{1}\chi_{1}  + { \rm h.c.} ~~ ~~~ \nonumber \\
\label{z4}
\end{eqnarray}
After symmetry breaking, the first term in Eq.~(\ref{z4}) leads to a mass splitting for the real and imaginary 
components $S,P$ of the field $\chi_{2}$, analogous to the $Z_2$ scalar model considered previously in Eq.~(\ref{z2S}), 
and  we take $\lambda_1$ negative so $m_S < m_P$. 
According to Eq.~(\ref{zn}) the theory has a descendant $Z_4$ discrete symmetry
under which both $\chi_2$ and $\chi_1$ are charged,
as well as a  $Z_2$ symmetry under which $S,P$ are even and $\chi_{1}$ is odd. The transformation properties are summarized in
Table~\ref{table}. 
It is clear that $\chi_{1}$, being the only particle with a nontrivial $Z_2$ charge, is stable and a DM candidate. 
However, if $S$ is lighter than twice the mass of $\chi_{1}$ it will also be stable by virtue of the $Z_4$ symmetry. 
Hence, in this case there are potentially two species of DM, though what contribution each 
makes to the cosmological DM depends on other considerations that we will examine later. 

As a slightly more complicated example, consider a $Z_6$ model with a charge 6 dark Higgs  $\phi_{6}$ and three scalar matter fields 
$\chi_{1}$, $\chi_2$, and $\chi_3$  with charges 1, 2, and 3. The Lagrangian contains
\begin{eqnarray}
\Delta{\cal L} & =& -\lambda_1 \phi^\dag_{6} \chi_{3} \chi_{3}   -\lambda_2 \phi^\dag_{6} \chi_{3} \chi_{2}\chi_1 
-\lambda_3 \phi^\dag_{6} \chi_{2} \chi_{2} \chi_2  \nonumber \\ &&
 - \lambda_4 \chi^\dag_{3} \chi_{2} \chi_{1} - \lambda_5 \chi^\dag_{3} \chi_{1} \chi_{1} \chi_{1} 
- \lambda_6 \chi^\dag_{2} \chi_{1}\chi_{1}  + { \rm h.c.} ~~ ~~~ \nonumber \\
\label{z6}
\end{eqnarray}
In this case, we can decompose $Z_6 \cong Z_3 \times Z_2$, so that there are two distinct cyclic symmetries. 
The transformation properties of the 
matter fields under these symmetries are displayed in Table~\ref{table}. We see again that there are potentially 
two species of DM. For example, if $\chi_1$ is the heaviest matter field, then $\chi_2$ and $\chi_3$ will
be stable due to their nontrivial charges under $Z_3$ and $Z_2$, respectively. Note that $\chi_3$ will be split into its
real and imaginary components due to dark symmetry breaking and the lightest of these will be stable.
\begin{center}
\begin{table}
\resizebox{3.3cm}{!}{
\renewcommand{\arraystretch}{1.5}
\begin{tabular}{| l || c | c | }
\hline
\multicolumn{3}{|c|}{$Z_4$ model} \\
\hline \hline
  &  $\chi_1$ &  $\chi_2$   \\
 \hline 
 $Z_4$  &  $\displaystyle{ e^{\frac{2\pi i}{4}\cdot 1}}$  &  $\displaystyle{e^{\frac{2\pi i}{4}\cdot 2}}$   \bigstrut \\ \hline
 $Z_2$ &   $\displaystyle{e^{\frac{2\pi i}{2}\cdot 1}}$ &  $1$  \\ \hline
  \end{tabular}}
$\quad  $
\resizebox{4.5cm}{!}{
\renewcommand{\arraystretch}{1.5}
  \begin{tabular}{| l || c | c | c| }
\hline
\multicolumn{4}{|c|}{$Z_6$ model} \\
\hline \hline
  &  $\chi_1$ &  $\chi_2$ &  $\chi_3$  \\
 \hline
 $Z_6$ &  $ e^{\frac{2\pi i}{6}\cdot1}$ &  $e^{\frac{2\pi i}{6}\cdot 2}$ &  $e^{\frac{2\pi i}{6}\cdot 3}$ \\ \hline
 $Z_3$ &   $e^{\frac{2\pi i}{3}\cdot 1}$ &  $e^{\frac{2\pi i}{3}\cdot 2}$ &  $1$ \\ \hline
 $Z_2$ &   $e^{\frac{2\pi i}{2}\cdot 1}$ &  $1$ &  $e^{\frac{2\pi i}{2} \cdot1}$ \\
    \hline
  \end{tabular}}
\caption{Transformation properties of matter fields in the $Z_4$ (left) and $Z_6$ (right) multi-field models.
Charges $q_i$ for a field $\chi_i$ are defined through their transformation under $Z_N$:
$\chi_i \rightarrow {\rm exp}(\frac{2 \pi i}{N}\cdot q_i)\chi_i$. }
\label{table}
\end{table}
\end{center}

From the first example based on the $Z_4$ model, we learn that a $Z_{p^{m}}$ symmetry with $p$ a prime number and $m$ a natural
number allows for potentially $m$ stable states, while the example with the $Z_6$ symmetry shows us that depending on $N$,
$Z_N$ may be decomposed into a direct product of smaller groups \cite{LUP}. 
We may surmise that for a general $ N= p_1^{m_1} p_2^{m_2} \cdots p_k^{m_k} $,  where $p_i$ is prime ($p_i \neq p_j$ for $i\neq j$) 
and $m_i$ is natural, $Z_N$ is decomposed into the product group
\begin{equation}
Z_N \cong Z_{p_1^{ m_1}} \times Z_{p_2^{ m_2}} \times \cdots \times Z_{p_k^{\! m_k}}.
\label{decompose}
\end{equation}
There will be at most $m_1 + m_2 +\cdots + m_k$ stable species for a given $Z_N$ symmetry, though the
actual number depends on the field content and the spectrum. A simple recipe to obtain the maximum number
of stable states is as follows: for each prime $p_i$ in (\ref{decompose}), add fields with U(1)$_D$ charges 
$Q = \{N/p_i,N/p_i^2, \cdots N/p_i^{m_i} \}$ and order the spectrum 
$m_{N/p_i} < p_i \, m_{N/p_i^2} < \dots < p_i^{m_i-1} \, m_{N/p_i^{m_i}}$. 

Besides multi-component DM, 
another important effect which can occur in multi-field $Z_N$ models is mass mixing induced by dark symmetry breaking. 
To illustrate, consider a model with a charge $N$ dark Higgs, and scalar matter fields  $\chi_{Q}$ and
$\chi_{Q+N}$ with charges $Q$ and $Q+N$. Gauge symmetry allows the following term:
\begin{eqnarray}
 \Delta{\cal L} = - \frac{\lambda}{\sqrt{2}}  \phi^\dag_{N} \chi^\dag_{Q} \chi_{Q+N} + {\rm h.c.} , 
\label{multimass}
\end{eqnarray}
 which induces a mass mixing between the matter fields $\chi_{Q}$ and
$\chi_{Q+N}$ following dark symmetry breaking. This can be diagonalized with an
orthogonal rotation leading to mass eigenstates $\chi_{a,b}$. Consider for simplicity the limit in which  
the bare masses are equal, $m^2_Q = m^2_{Q+N} \equiv m^2$, and a small mass splitting induced by the
operator in (\ref{multimass}) given by $\Delta m^2 = \lambda v \ll m $. The mixing angle in this case is 
$45^{\rm o}$, and the masses are split by a small amount $\delta \sim \Delta m^2/M$.   
Besides a diagonal gauge interaction between the charged matter fields, 
this rotation induces an off-diagonal gauge interaction: 
\begin{equation}
i g_D V_\mu \left( \chi^\dag_a ~ \chi^\dag_b \right)
\left( \begin{array}{cc}
       Q+N/2 &  N/2 \\
  N/2  & Q + N/2
       \end{array}
\right)\overleftrightarrow {\partial_\mu}
\left( \begin{array}{c}
         \chi_a \\ \chi_b
       \end{array}
 \right).
\label{Voff}
\end{equation}
These interactions can lead to important consequences for direct detection experiments, 
as DM may undergo both elastic and inelastic scattering with nuclei, leading to a distinctive recoil spectrum. 
Moreover, off-diagonal terms allow for novel decay modes of $Z_N$ partners at colliders. We will explore these 
issues below.

One final common occurrence in $Z_N$ models is the appearance of nontrivial interactions between matter particles. 
Cubic and quartic vertices are allowed by gauge invariance so long as the corresponding charges cancel, and these 
can lead to important effects in DM physics. For example, one may have a 
Yukawa-type interaction between scalars $\chi_i$ and fermion matter fields $\psi_i$,
\begin{equation}
 \Delta {\cal L} \supset -\lambda \chi_i \psi_j \psi_k + {\rm h.c.}, 
\label{yuk}
\end{equation}
so long as the charges satisfy $i+j+k = 0$.
This interaction, along with similar cubic and quartic scalar matter interactions, can manifest in DM annihilation, 
affecting the relic abundance and indirect detection prospects, and lead to 
cascade decays of $Z_N$ partners given a production mechanism at collider experiments.


{\it Chiral models.} Lastly, we will consider models in which matter consists of a chiral set of fermions. This means the  
fermion content is chosen so that 1) there are no gauge anomalies, and 2)  a vectorlike mass term is forbidden by gauge symmetry. 
Indeed, it is a striking fact that each generation of SM matter  
forms a chiral set given the gauge group SU(3)$_c\times$SU(2)$_L\times$U(1)$_Y$, and so 
it seems quite plausible that DM could also be chiral.

Given that the dark fermions are charged only under U(1)$_D$, there are two 
anomaly conditions to satisfy: the U(1)$_D^3$ anomaly and the U(1)$_D$-gravitational anomaly. 
To forbid mass terms, any two charges must not sum to zero. For fermions with charges 
$Q_{i}$, these conditions are simply
\begin{eqnarray}
\sum_{i} Q_{i}^3 &=& 0, \nonumber \\ 
\sum_{i} Q_{i} &=& 0,  \nonumber \\
Q_{i}+Q_{j} &\neq &  0, \quad {\rm for~all}~ i,j ~.
\label{chiral}
\end{eqnarray}
Ref.~\cite{anomaly}  analyzed quite generally the problem of finding fermion charges satisfying the conditions (\ref{chiral}) 
and presented many examples of chiral sets. While there are clearly 
a vast number of possibilities for models of chiral DM, we will be content here to illustrate a few common features
in a simple model. Consider the following chiral set:
\begin{equation}
 \psi^{i}_{-1}, \psi_{3}, \psi_{4}, \psi^{i}_{-6}, \psi_{7},
\label{chiralset}
\end{equation}
where $i=1,2$. This set contains 7 `flavors' of Weyl fermions, and it is straightforward to check that the conditions (\ref{chiral}) 
are satisfied. Besides the fermions in the chiral set, we will need at least one dark Higgs to break the gauge symmetry and to 
generate masses for the fermions. We choose a charge 7 Higgs field $\phi_{7}$ to accomplish this task. This means that 
ultimately there will be a discrete $Z_7$ symmetry that stabilizes the lightest fermion mass eigenstate, but as we will see 
there may also be accidental global flavor symmetries which can stabilize multiple candidates. 

The Higgs $\phi_{7}$  allows for $\psi_{3}, \psi_{4}$ to marry through the 
Yukawa operator $\phi^\dag_{7}\psi_{3} \psi_{4} + {\rm h.c.}$, to form 
a Dirac fermion. Furthermore, $\psi^i_{-1}, \psi^j_{-6}$ also are married by the Higgs through the 
following Yukawa interaction:
\begin{eqnarray}
\Delta {\cal L} = - \lambda^{ij} \phi_{7} \psi^i_{-1} \psi^j_{-6} + {\rm h.c.} ,
\end{eqnarray}
where $\lambda^{ij}$ is a general matrix. This Yukawa mixing is reminiscent of what occurs in the 
SM, but we should remember that these fermions are not part of separate `generations'. The Yukawa matrix can be diagonalized
by separate unitary transformations on  $\psi^i_{-1}$ and  $\psi^j_{-6}$ and these unitary matrices are not observable with 
the field content we have so far specified. In particular,  there will  be no dark flavor change induced 
by the gauge interactions, as there is an analogue of the GIM mechanism operating in this model. 

What about the remaining fermion $\psi_{7}$ in the chiral set (\ref{chiralset})?  This fermion is quite similar to the neutrinos
of the SM as no tree-level mass may be generated with the matter and Higgs content considered so far. How may we generate
mass for this fermion? First, let us form the U(1)$_D$ neutral operator $\phi^\dag_{7} \psi_{7}$ having dimension 5/2 and 
spinor Lorentz index. This operator, being analogous to $HL$ of the SM, immediately suggests mechanisms for mass generation. A
Majorana mass term may arise from the dimension five operator  $(\phi^\dag_{7} \psi_{7})^2$, or we may add a singlet fermion
$\xi_{0}$ and form a Dirac mass  $\phi^\dag_{7} \psi_{7}\xi_{0} + {\rm h.c.}$. As the singlet may also have a bare mass, a 
see-saw mechanism is potentially operative. If U(1)$_D$ is broken around the weak scale, and the dimension 5 Majorana mass operator
or the singlet mass is generated at a higher scale, then 
typically we are led to predict a much lighter fermionic state in the spectrum. This state is potentially stable and 
may be difficult to deplete in the early universe, having few annihilation channels which are unsuppressed. There are many 
ways out of this. For example, the singlet $\xi_{0}$  may couple to the SM neutrinos, breaking the accidental flavor symmetry 
and allowing the state to decay. We may add new dark Higgs fields to make the neutrino-like state heavy, 
one example being a charge 14 Higgs which would not disturb the subgroup discrete $Z_7$ symmetry.  Alternatively, 
a very light or massless state in the dark sector could 
manifest itself as a new component to the cosmological dark radiation and have interesting signatures in its own right.

Finally, let us elaborate on the symmetries of this model. By construction, there is a $Z_7$ discrete symmetry which
will forbid the lightest state with nontrivial charge from decaying. However, there are additional accidental flavor symmetries 
present which lead to multiple stable states. Let us trace the origin of these symmetries. 
First, if we turn off all interactions, there is a large U(7) flavor symmetry, which is broken to  
U(2)$^2 \times$U(1)$^3$ by the gauge
interactions. The Yukawa terms further break the flavor symmetry to U(1)$^4$ leading to potentially 4 stable mass eigenstates.
The U(1) flavor symmetry associated with the neutrino-like state $\psi_{7}$ may be broken further to a $Z_2$ symmetry
by the inclusion of a dimension 5 Majorana mass operator, or to a diagonal subgroup of the combined U(1)$^2$ symmetry with 
the inclusion of the singlet $\xi_0$. Finally, higher dimension operators will generically further break these accidental 
symmetries. For example, integrating out a charged 2 scalar field (which does not condense) generates operators like 
$(\psi_{-1}\psi_{-1})(\psi_{-1}\psi_{3})$, $(\psi_{4}\psi_{-6})(\overline{\psi}_{-1}\overline{\psi}_{-1})$, etc., 
allowing some of the would-be stable states to decay. However, the discrete $Z_7$ symmetry is exactly conserved and will ensure
the stability of the lightest charged state, thus providing a DM candidate.


{\bf Phenomenology.} We will now describe basic aspects of the phenomenology in this class of models. Detailed
scans of the parameter spaces of specific models is beyond the scope of the present work. Rather, we will focus
on the novel implications of a larger $Z_N$ discrete symmetry and
the impact of additional states in the dark sector such as $Z_N$ matter as well as dark gauge and Higgs bosons. 
The first question we must address is the following: how can the dark sector communicate
with the SM? 


{\it Portals.} Given that we are considering models in which DM is in a hidden sector (DM carries no charge 
under the SM gauge group, and likewise the SM fields carry no charge under U(1)$_D$), if interactions are to exist
we expect that at high scales there exist heavy states charged under both sectors. Integrating these out, we generate 
all higher-dimensional operators allowed by the gauge symmetries of the theory. We may also generate `portal' 
operators~\cite{portal}: renormalizable, gauge-invariant operators that connect the SM to the dark sector. 
In particular, a dark sector based on a broken U(1)$_D$ gauge symmetry affords 
the opportunity to interact through both the kinetic mixing~\cite{holdom} and Higgs portal~\cite{singlet1}:
 \begin{eqnarray}
{\cal L}_{portal} &=& 
-\frac{\kappa}{2} V_{\mu\nu} B^{\mu\nu}  
-2 \lambda_{\phi}\phi^\dag \phi H^\dag H -2 \lambda_{\chi} \chi^\dag\chi H^\dag H,  \nonumber \\
\label{portal}
\end{eqnarray}
where of course the final term is present only for scalar matter fields at the renormalizable level. 
Note that the combination of vector and Higgs portals was first studied in Ref.~\cite{foot}.
Besides kinetic mixing between vectors, symmetry breaking induces mass mixing between the SM and dark 
Higgs bosons. We will limit ourselves to the case of small mixing arising from portals, so that it suffices to 
treat the terms in Eq.~(\ref{portal}) as interactions. 

Alternatively, one can diagonalize the Lagrangian and work directly with the physical states. 
The  main effect of the kinetic mixing is that  
SM matter picks up a small charge under U(1)$_D$ proportional to the kinetic mixing strength $\kappa$:
\begin{equation}
 {\cal L} \supset  \kappa V_\mu[-c_w J_{EM}^\mu + s_w (1-m_Z^2/m_V^2)^{-1}J_Z^\mu].
\label{VSM}
\end{equation}
The Higgs portal mass mixing may be treated by first 
expanding around the vacuum, $v\rightarrow v+h$ and $v'\rightarrow v'+h'$, followed by diagonalizing the
system with an orthogonal rotation by angle 
${\theta}_h \simeq 2 \lambda_\phi v v' / m_h^2$, valid in the limit of $m_{h'} \ll m_h$.
Diagonalization induces interactions between the SM Higgs and the dark sector, as well as the dark Higgs with the SM,
\begin{equation}
 {\cal L} \supset \theta_h ( h' J_h - h J_{h'}),
\end{equation}
where $J_{h,h'}$ are the SM and dark currents coupling to the Higgs bosons. Note that for a dark 
symmetry breaking scale well below the weak scale, there can be an issue with technical 
naturalness since the SM Higgs vev will tend to raise the dark Higgs vev for large Higgs portal 
coupling $\lambda_\phi$ in Eq.~(\ref{portal}). In some cases we will be interested in couplings of order 
the bottom quark Yukawa, so that e.g. a symmetry breaking scale in the dark sector of $v'\sim 10$ GeV
requires only a mild order one cancellation between different terms in the scalar potential. Scales much lighter 
than this require smaller values of $\lambda_\phi$ for technical naturalness. Similar comments apply to the mass 
of scalar DM when the Higgs portal coupling $\lambda_{\chi}$ in Eq.~(\ref{portal}) is sizable.


{\it Existing constraints.} Since the dark sector carries no SM charges, new states 
can in principle be much lighter than the weak scale and still avoid 
constraints coming from direct production and precision measurements. 
The constraints that do exist depend
on the strength of the portals and the precise decay patterns of dark states. We now 
give an overview of the possible constraints for various mass scales in the dark sector.

Let us first discuss constraints on the kinetic mixing portal. If the 
dark gauge boson mass is above the weak scale, there are mild 
constraints on the kinetic mixing parameter $\kappa$ 
coming from electroweak precision tests,
$\kappa \la 10^{-1} - 10^{-2}$~\cite{ewpt}. Vectors with masses 
$O(1\,{\rm GeV}) < m_V < O(100\,{\rm GeV})$ can be constrained further from a variety of $e^+ e^-$ collider 
data, and give a rough bound of $\kappa \la {\rm few} \times 10^{-2}$~\cite{wacker}. 
For sub-GeV vectors, additional constraints arise from corrections to the anomalous 
magnetic moment of the electron and muon, extending $\kappa$ down to 
$\la {\rm few} \times 10^{-3}$ \cite{maxim}. There are also 
model-dependent constraints for light vectors, assuming the
dominant decay mode of $V$ occurs through the 
kinetic mixing portal back to SM states, arising from searches at $B$ and other
meson factories as well as fixed target experiments~\cite{Bfac,babar,slacfixed,BPRfixed}. 
In particular the fixed target experiments are able to exclude a large portion of the parameter 
space for vector masses below a few hundred MeV, with constraints on 
$\kappa$ ranging from $10^{-4} - 10^{-8}$~\cite{slacfixed,BPRfixed}. 
There has been a great deal of effort invested in the last two years 
to constrain or search for new light gauge bosons
decaying to leptons, primarily motivated by a possible DM 
connection \cite{darkforce} to cosmic ray anomalies \cite{pamela,fermi}. There are the 
additional possibilities of detecting or constraining light gauge bosons 
decaying to leptons through searches of the 
existing $B$-factory data sets~\cite{Bfac} and new fixed target experiments~\cite{slacfixed}.

Constraints on the Higgs portal are much weaker and only model-dependent bounds are possible.
Very large $O(1)$ mixing angles $\theta_h$ can affect electroweak precision observables~\cite{Barger}, 
but for a sub-weak scale dark Higgs $h'$, a stronger constraint comes from direct production 
at the LEP experiments. The mixing induces a coupling of the dark Higgs to the $Z$ boson,  allowing for a dark Higgs-strahlung 
process $e^+ e^- \rightarrow Z h'$. Data from the LEP experiments 
constrain the parameter $\xi^2 \equiv (g_{\rm HZZ}/g_{\rm HZZ}^{\rm SM})^2 \la 10^{-2}$ \cite{lep} for a very light 
Higgs-like state, which can be translated to a bound on the mixing angle $\theta \la 0.1$ which is easily satisfied for 
$\lambda_\phi \sim y_b$. Note that this bound is conservative and could in principle be weakened considerably 
in specific models depending on the decay modes of $h'$. For very light states in the GeV range, Higgs portal couplings
can be probed by rare heavy-flavor meson decays \cite{Bdecay}, potentially down to mixing angles of $10^{-2}-10^{-3}$, 
though again, these constraints depend on the decays of the GeV-scale dark states.


{\it Relic Abundance.}  
Assuming that annihilation processes dominate at freezeout, 
the relic abundance of a stable heavy particle is given by 
$\Omega_{DM}h^2  \simeq  0.1 \times (\langle \sigma v \rangle/{\rm pb} )^{-1}$,
where $\langle \sigma v \rangle$ is the DM annihilation cross section.
The predicted relic abundance is to be compared with the observed value from WMAP, 
$\Omega_{DM}h^2 = 0.1123\pm 0.0035$ \cite{wmap}.
We may classify the annihilation channels according to 
whether the final states belong to the dark sector or the SM:
\begin{equation}
 \begin{array}{rl}
 {\bf \rm I}: & \chi + \overline{\chi}   \rightarrow  V V, V h', h'h', \dots \\
&\\
{\bf \rm  II}: &  \chi + \overline{\chi}  \rightarrow  V^*, h'^*, h^* \rightarrow \psi_{SM} + \overline{\psi}_{SM}   
\end{array}
\nonumber
\end{equation}
Case  {\bf \rm I}  corresponds to the `secluded' DM regime~\cite{PRV}, in which the relic abundance does not 
depend on  the strength of mediation to the 
SM. The secluded regime is typically most important when the DM candidate(s) is not the lightest state in the dark sector and 
direct annihilation channels are kinematically open in the dark sector. 
For example, if $m_V < m_{\chi}$ the annihilation cross section is parametrically
$\langle \sigma v\rangle_{\chi\overline{\chi}\rightarrow VV} \approx \pi \alpha_D^2/m_{\chi}^2$, 
so that the observed WMAP abundance  requires $\alpha_D \ga 5\times 10^{-3}  (m_{\chi}/200 \,{\rm GeV})$~\cite{PRV}.

If no annihilation channels are available directly in the dark sector, then it is still possible to obtain the proper 
relic abundance provided the portal operators are active and have a sizable coupling. This is case {\bf \rm II} above 
in which DM may annihilate through
the kinetic mixing or Higgs portal operators into the light SM final states.  
As an example, consider the $Z_2$ scalar model of Eq.~(\ref{z2S}). If the DM candidate $S$ is the 
lightest state in the dark sector, it must annihilate into SM states if it is to be thermally depleted. 
One option is to utilize the `direct' Higgs portal operator 
in Eq.~(\ref{portal}), $(H^\dag H)(\chi^\dag \chi) \supset h SS$. In the regime $m_S \ll m_h$ and away from the 
$s$-channel resonance at $m_S = m_h/2$, the dominant annihilation mode is into $b\bar{b}$. The annihilation cross section is
$\langle \sigma v\rangle_{SS\rightarrow b\bar{b}} \simeq 3\lambda_\chi^2 m_b^2/\pi m_h^4$, which dictates that 
$\lambda_\chi \ga 0.2\times (m_h/120\,{\rm GeV})^2$. This is a sizable coupling, much larger than the bottom quark Yukawa,
and so this model makes the interesting prediction that the SM Higgs decays to DM. 
In fact in the limit that all other states in the dark sector 
are heavy, this model matches on to the minimal $Z_2$ scalar singlet model of Refs.~\cite{singlet1,singlet2,singlet3}. 
 
Next, we consider the relic abundance in multi-component models. Typically we might expect that if two DM 
particles have distinct annihilation channels governed by different couplings and mass scales, then barring some 
accidental coincidence in these couplings, the two DM states will end up with dramatically different relic densities
and one state will dominantly make up the observed cosmological DM. The situation is somewhat more constrained
if the two states have the same annihilation modes and symmetry guarantees the equality of couplings. This occurs, for example,
when two heavy states annihilate into light vectors. Given our discussion above regarding this channel, we expect then if 
two states $\chi_i$ , with $i=1,2$,  have charges $Q_i$ and masses $m_i$, their relative abundance will be given by 
$\Omega_1 / \Omega_2 =  (m_1^2/m_2^2)(Q_2^4/Q_1^4)$. In this case, a similar relic abundance requires a coincidence of masses
and charges. However, even if a particular species is subdominant in the cosmic DM, it still may 
give signatures in direct and indirect detection experiments. We will illustrate this below for multi-component 
direct detection.

The final interesting possibility we wish to mention is the process of semi-annihilation, examined recently in Ref.~\cite{semiA}.
Typical annihilation processes change the number of DM particles by two units. In contrast semi-annihilation reactions
change the number by only one unit. The simplest example of this occurs in the $Z_3$ scalar model in 
Eq.~(\ref{z3})~\cite{semiA}. If the 
dark Higgs $h'$  is lighter than the DM $\chi_1$, then the process $\chi \chi \rightarrow \chi h'$ will occur with an annihilation 
cross section $\langle \sigma v\rangle_{\chi\chi\rightarrow \chi h'} = 3 \lambda^2 / 128 \pi m_\chi^2$, implying that
$\lambda \ga 0.1 \times (m_\chi/200\, {\rm  GeV})$ if other annihilation processes are small. In fact semi-annihilation 
occurs quite generally in multi-component DM models, where processes such as $\chi_i \chi_j \rightarrow \chi_k h'$, etc, 
can take place, and we refer the reader to Ref.~\cite{semiA} for more details.


{\it Direct Detection.}  We now discuss the possibility of direct detection of DM. Scattering with nuclei is mediated via a 
$t$-channel exchange of vectors or Higgs bosons through  the  portal 
interactions in (\ref{portal}). Vector exchange leads to the following 
DM-nucleon scattering cross section~\cite{PRV}:
\begin{equation}
 \sigma_n \simeq \frac{16 \pi  \kappa^2 c_w^2 \alpha \alpha_D Q^2 \mu_n^2}{m_V^4}\frac{Z^2}{A^2}.
\label{signV}
\end{equation}
The most recent limits from the CDMS-II~\cite{cdms} and XENON100~\cite{xenon} experiments have 
reached a sensitivity of $\sigma_n \la (3\!-\!4) \times 10^{-44}$ cm$^2$ for DM with masses larger than
around 50 GeV \cite{otherkinetic}. This translates to a bound $\kappa g_D Q \la 10^{-5} \times(m_V/10 \, {\rm GeV})^4 $ for the vector portal. 

Exchange of the SM Higgs boson due the portal interaction also mediates elastic scattering with nuclei. As the simplest example
consider the $Z_2$ scalar model in Eq.~(\ref{z2S}). The $S$-nucleon cross section is (see e.g. \cite{singlet3})
\begin{equation}
  \sigma_n \simeq \frac{\lambda_\chi^2 f^2 m_n^4}{\pi m_\chi^2 m_h^4},
\end{equation}
where $f \equiv \sum  \langle n | m_q \overline{q} q | n \rangle / m_n \simeq 0.35$  for where $n$ is a proton or neutron 
\cite{savage}. Constraints from CDMS-II~\cite{cdms} and XENON100~\cite{xenon} imply 
$\lambda_\chi \la 0.03 \times (m_\chi/ 50\, {\rm GeV}) (m_h/120\,{\rm  GeV})^2$. Note that for smaller masses in the 10 GeV range, 
the sensitivity of these experiments falls off rapidly and larger values of the portal couplings can 
be accommodated. Indeed, there has been some excitement for DM with masses
in the 10 GeV range, since such a light DM candidate has the potential to explain the 
longstanding DAMA/NAI and DAMA/LIBRA \cite{dama} signal as well as the recent events
from COGENT \cite{cogent}. This framework can accommodate such a light DM candidate as well, but we will
not pursue this possibility further here. 

Another possibility to explain the DAMA results is inelastic DM~\cite{inelastic}. For a recent study of the status of inelastic DM, 
see Ref.~{\cite{poker}}. 
As we have seen, mass splittings may occur 
very easily in particular models with discrete gauge symmetries. For example, the models with a $Z_2$ discrete symmetry naturally
lead to inelastic scattering due to the off-diagonal coupling to the vector $V$ in Eqs.~(\ref{VPS},\ref{V12}), and in fact, very 
similar models were originally considered as examples in \cite{inelastic}. More generally, 
inelastic couplings occur for a multi-field $Z_N$ model when there exists matter fields $\chi_{N/2}$ with charge $N/2$, 
or two matter fields with charges $Q$ and $Q+N$ as in 
Eq.~(\ref{multimass}). Small mass splittings of $O(100\, {\rm keV})$ can lead to a very long-lived excited state with
a significant relic abundance, as pointed out in Ref.~\cite{exo1}, and this leads to the possibility of exothermic 
down-scattering with nuclei. Recently, this has been utilized in Refs.~\cite{exo2,exo3} as a possible explanation of DAMA.

There is also the possibility of `semi-elastic' scattering, 
investigated recently in Ref.~\cite{semielastic}, which has a very distinctive recoil spectrum. Semi-elastic scattering
can occur due to the exchange of a vector with an off-diagonal coupling and a Higgs with a diagonal coupling, as in 
\cite{semielastic}, or through the exchange of a vector with both diagonal and off-diagonal couplings as in Eq.~(\ref{Voff}). 
In the latter case, in order to have the small mass splitting required for an inelastic transition, 
the bare masses of the two fields must be approximately equal, and we have not provided a symmetry reason for 
this here though in principle this could be addressed with further model building. 

\begin{figure}
\centerline{
\includegraphics[width=0.4\textwidth]{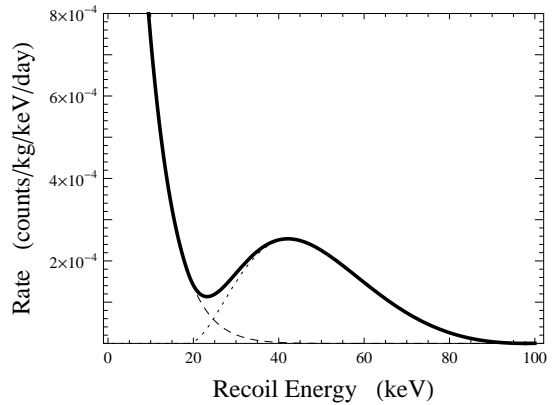}}
\caption{Recoil spectrum of two-component scattering at XENON100. Here we consider a $Z_6$ model in Eq.~(\ref{z6}) 
with two DM particles $S \supset  \chi_3$  and $\chi_2$. 
We have assumed masses for the DM particles $m_S = 3$ TeV with a splitting of $\delta \equiv m_P-m_S = 150$ 
keV leading to inelastic transitions,
and $m_{\chi_2} = 30$ GeV. Other parameters entering into the relic abundance and scattering cross section are 
$\kappa  = 8 \times 10^{-4}$, $\alpha_D\sim 0.01$
and $m_V= 10 $ GeV. This point is consistent with current direct detection constraints \cite{cdms,xenon} and should 
lead to $\sim$ 15 events at XENON100 assuming an effective exposure of 3000 kg-days. 
The dashed(dotted) lines show the elastic(inelastic) components while the solid line 
shows the total rate for summer (June 2nd). To compute the rate, we have used a standard Maxwellian velocity distribution
with $v_{esc}=500$ km/s. }
\label{multi}
\end{figure} 

Finally, we turn to the direct detection signatures of multi-component DM, which has been considered previously in Ref.~\cite{dual}. 
Naively, one might expect that detecting two components is  only feasible if both DM particles have similar relic densities. As we
discussed above, this is not generally true unless the both particles have similar masses and annihilation channels. 
However, the rate depends both on the relic density and the scattering cross section. A particle with 
a large annihilation cross section that is efficiently depleted early in the universe 
might also be expected to have a larger scattering cross section with nuclei. Furthermore, 
the particle with the dominant relic density can scatter inelastically with nuclei, which leads to a smaller 
cross section compared to the elastic case. These two facts suggest that it may indeed be possible
to have significant rates for both DM particles at direct detection experiments. 

As an example, consider the $Z_6$ model in Eq.~(\ref{z6}) with two 
stable particles, the real component $S$ of $\chi_3$ and $\chi_2$. 
These states are stable due to the presence 
of $Z_3$ and $Z_2$ symmetries as explained above.  
Assuming annihilations to dark gauge bosons $V$, whichever particle is heavier 
will dominantly comprise the cosmological DM, and we assume this to be $S$.
Because the $S$ is split from its partner $P$, an off-diagonal
gauge interaction as in Eq.~(\ref{VPS}) mediates inelastic scattering with nuclei, and this will be quite suppressed
for splittings of $O(100\,{\rm keV})$. 
On the other  hand $\chi_2$ will undergo elastic transitions, and because of its dilute presence in the galactic halo,
we are free to increase $\kappa$ to much larger values without conflicting with direct detection constraints.
If there is a large mass separation between the two DM components, a very distinctive recoil spectrum can
be observed at current and next generation experiments, as we illustrate in Fig.~\ref{multi}. 


{\it Collider Signatures.}
When considering new particles from a dark sector, because of the weak coupling to the SM it is typically hopeless to directly
produce such states at colliders. However, an interesting opportunity arises if a hidden valley scenario is operative~\cite{HV}, 
in which case we first produce new heavy `connector' states with SM quantum numbers in abundance, 
which subsequently decay into the dark sector states. While there are many possibilities for connector states around or above the 
weak scale, we will focus on the minimal case of the SM Higgs boson as our connector. 
The SM Higgs is in fact an ideal choice for a bridge to the dark sector because of its sizable production cross section, 
$\sigma_{gg\rightarrow h}^{NNLO}\sim 50(10)$ pb for LHC at $\sqrt{s}=14(7)$ TeV, and narrow width. For $m_h<130$ GeV, 
the Higgs decay width is governed by the small bottom quark Yukawa $y_b \sim 1/60$, so that a light SM Higgs is quite 
susceptible to new decay modes~\cite{higgsdecay}. 
We will utilize this property of the Higgs to gain access to states in the dark sector.

From Eq.~(\ref{portal}) we see there are two possible operators that will allow for new decays of the SM Higgs. 
The operator $\phi^\dag \phi H^\dag H$ allows  for the decay of the SM Higgs to dark Higgs and gauge bosons, with partial 
widths
\begin{eqnarray}
 \Gamma_{h\rightarrow h'h'} & = & \frac{\lambda_{\phi}^2 v^2}{8 \pi m_h}\left(1-4 x_{h'}^2\right)^{1/2}, ~~~~~~\\
 \Gamma_{h\rightarrow VV}   & = & \frac{\lambda_{\phi}^2 v^2}{8 \pi m_h}\left(1-4 x_V^2\right)^{1/2} 
\left(1-4x_V^2 + 12 x_V^4\right), ~~~~~~
\end{eqnarray}
where $x_i\equiv m_i/m_h$. The SM Higgs couples to the dark vectors due to the mass mixing induced by ${\cal L}_{portal}$.
Notice that in the limit $x_{h',V} \ll 1$, the branching to dark Higgs bosons and dark vectors is equal, 
$\Gamma_{h\rightarrow h'h'} = \Gamma_{h\rightarrow VV} \simeq \lambda_{\phi}^2 v^2 /8 \pi m_h $, a simple consequence of the
Goldstone-equivalence theorem since the Higgs portal (\ref{portal}) leads to identical couplings of the SM Higgs 
to the radial and Goldstone modes of the dark Higgs. We should compare these partial widths with the partial width to a pair of 
bottom quarks, $\Gamma_{h\rightarrow b\bar{b}} \simeq 3 y_b^2 m_h/16 \pi$, telling us that for $\lambda_\phi \sim y_b$ 
the branching into 
dark Higgs and gauge bosons will be significant. Note that this value of $\lambda_\phi$ corresponds to a mixing angle 
$\theta_h \sim 10^{-3}-10^{-2}$, well below any existing constraints. 

\begin{figure}
\centerline{
\includegraphics[width=0.43\textwidth]{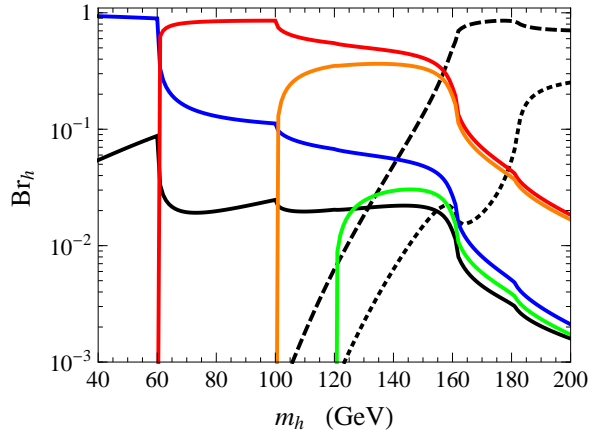}}
\caption{SM Higgs branching ratios. Here we are considering the $Z_2$ scalar model of Eq.~(\ref{z2S}), with a spectrum
$(m_V,m_S,m_P,m_{h'} )= (20,30,50,60 )$ GeV, and their corresponding branching ratios are indicated by 
$(VV,SS,PP,h'h')= ({\rm blue, red, orange, green} )$. We have taken portal couplings to be $\lambda_\phi\sim y_b = 1/60$, and 
$\lambda_\chi = 1/20$. The SM $b\bar{b}, ZZ$ and $WW$ modes are indicated  by the (solid, dotted, dashed) black lines. Below the 
$WW$ threshold, the SM Higgs decays dominantly to hidden sector states.}
\label{branch}
\end{figure}

If the theory contains scalar matter fields $\chi$, where $\chi$ now generically denotes
either the stable DM state or an unstable $Z_N$ partner, then the second Higgs portal operator 
$\chi^\dag \chi H^\dag H$ in Eq.~(\ref{portal})
allows a  decay $h \rightarrow \chi^\dag \chi$, with partial width
\begin{equation}
 \Gamma_{h\rightarrow \chi^\dag \chi} = \frac{\lambda_{\chi}^2 v^2}{4 \pi m_h}\left(1-4 x_{\chi}^2\right)^{1/2}. ~~~~~~
\end{equation}
A similar conclusion applies: as long as $\lambda_\chi$ is comparable to $y_b$, the SM Higgs will have a sizable branching 
into DM or its $Z_N$ partners. We show the branching ratios of the SM Higgs in Fig.~\ref{branch} for a sample model and spectrum.

After the primary decay of the SM Higgs into either $h'h', VV$, or $\chi^\dag\chi$, the observed signature
depends on the cascade decays of the dark states. This becomes highly dependent on both the field content and
the spectrum in the dark sector. Even a small number of additional unstable $Z_N$ matter states in the dark sector 
can lead to very complicated signatures due to long cascades, 
particularly when there is mass mixing or nontrivial $Z_N$ interactions 
as in (\ref{yuk}). It is therefore useful to first consider a simplified situation where there is one stable DM
$\chi$ and a heavier unstable $Z_N$ partner $\chi^*$, in which case 
we can classify the possible decay patterns as follows:
\begin{itemize}
\item $\chi$ stable ($\not \!\!E )$,
\item $\chi^*  \rightarrow  2\chi , V \chi,  h \chi$ ,
\item $V \rightarrow  2\chi, 2\chi^*, \chi\chi^*, \ell^+ \ell^-$, $jj$ ,
\item $h'\rightarrow 2\chi, 2\chi^*, \chi\chi^*, VV, jj$ . 
\end{itemize}
Note that this is schematic, since for the matter states $\chi$, $\chi^*$ we have not distinguished between particle and antiparticle,
and furthermore in particular models the real and imaginary components of matter fields may receive mass splittings. 
We observe two important points from the decay patterns above: 
1) It is quite common to decay into a stable DM state, which will lead to a missing 
energy signature, and 2) dark states may cascade decay back to the SM, leading to a dilepton or dijet signature. 
Thus the generic signatures in this class of models are multi-lepton, multi-jet final states with missing energy.

While the origin of the missing energy signature is fairly obvious, it is useful to examine in more detail 
how it is possible for dark sector states to decay back to the SM. This occurs due to the 
kinetic mixing and Higgs portals in Eq.~(\ref{portal}). For the vector, the kinetic mixing operator 
$V_{\mu\nu} B^{\mu\nu}$ leads to the coupling (\ref{VSM}) of $V$ to the SM fermions. 
The partial width for a vector to decay to a generic SM fermion $f$ is
$\Gamma_{V \rightarrow f \bar f} \simeq (N_c/3) \alpha \kappa^2 m_V$.
If there are no direct channels open to dark sector states, the vector
will decay to SM fermions. For very light vectors, below the hadronic threshold, 
the vectors decay into $e^+e^-$ pairs or, if kinematically allowed, $\mu^+\mu^-$ pairs. 
Above the dipion threshold, the vector can decay to hadrons; the hadronic yield is governed by 
the well-measured form factor $R=\sigma_{ e^+e^-\rightarrow hadrons}/\sigma_{ e^+e^-\rightarrow \mu^+\mu^-}$ for vectors
lighter than several GeV~\cite{Bfac,pdg}. At higher masses, 
one can use the quark description to calculate the decay to hadrons~\cite{h4f}.
In summary, the vector decays fairly democratically, and in particular there is always a significant $O(1)$ 
branching to dilepton final states. Finally, the decays are typically prompt for $\kappa \ga 10^{-5}$, but there 
may be displaced vertices for smaller kinetic mixing parameters. Similarly, the dark Higgs $h'$ can decay via 
the Higgs portal to the heaviest kinematically available SM fermion pair. 
Typically this is the $b$-quark pair, and so the partial decay is given by  
$\Gamma_{h'\rightarrow b\bar{b}} \simeq  3 \theta_h^2 y_b^2 m_h/16 \pi$, which is a prompt decay for $\theta_h \ga 10^{-4}$.
A leptonic signature from the Higgs portal is not possible unless the dark Higgs is very light, in 
the hundreds of MeV range, in which case there is some tension with technical naturalness. From now on we will focus
on the clean leptonic signature via the kinetic mixing portal. The possibility of the SM Higgs decaying to multi-lepton final
states in this manner was discussed in Ref.~\cite{h4f}. 

Because of the kinetic mixing suppression in the decay $V\rightarrow f \bar{f}$, if kinematically allowed the dark vector 
will prefer to decay directly into states in the dark sector. However, this does not necessarily imply an absence of 
SM particles in the final states, since other dark sector states may decay back to the SM through an off-shell $V^*$.
This occurs when mass splittings or mixings lead to off-diagonal gauge interactions, as given e.g. in 
Eqs.~(\ref{VPS},\ref{V12},\ref{Voff}). As an illustration, consider the $Z_2$ scalar model with the 
gauge interaction in Eq.~(\ref{VPS}),
and a spectrum such that $m_V \gg m_P > m_S$, so that on-shell vectors decay dominantly via $V \rightarrow PS$. 
The heavy scalar partner will decay via the three-body process $P \rightarrow
S V^* \rightarrow S f \bar{f}$, and in the limit of large $m_V$ and small fermions mass the 
partial width is given by
\begin{eqnarray}
 \Gamma_{P\rightarrow S\overline{f}f}&=& \frac{N_c \, \alpha \alpha_D  \kappa^2 (g_V^2 +g_A^2)}{48 \pi \, c_w^2}\frac{m_P^5}{m_V^4}
 f\left( \frac{m_S^2}{m_P^2 } \right), ~~~~~~
\end{eqnarray}
where $f(\hat{x})=1 - 8 \hat{x} - 12 \hat{x}^2 \log{\hat{x}} +8 \hat{x}^3 -\hat{x}^4$. The quantity 
$(g_V^2+g_A^2)$ can be extracted from Eq.~(\ref{VSM}) and depends on the ratio
$m_V/m_Z$; light vectors $m_V/m_Z\ll1$ couple to electric charge, $(g_V^2+g_A^2) \simeq c_w^4 Q_{EM}^2$,  while
heavy vectors $m_V/m_Z\gg 1$ couple to hypercharge,  $(g_V^2+g_A^2)\simeq (Y_L^2 +Y_R^2)/2$. For a particular 
fermion final state to be allowed kinematically, the mass splitting must be large enough, $m_P-m_S > 2 m_f$. 
The lifetime depends  sensitively on the masses of the vector and $Z_2$ partner $P$ as well as
the kinetic mixing, but for masses in the 10-100 GeV range the displaced vertices may occur even for 
large values of the kinetic mxing $\kappa\la 10^{-3}$.

An interesting probe of multi-component $Z_N$ models would be the presence of two or more 
distinct stable particles in the cascade decays of the SM Higgs. This may happen from an initial asymmetric 
decay of the SM Higgs into two distinct states, $h\rightarrow \chi_i \chi_j$ with $i\neq j$ which subsequently
cascade into distinct stable states plus SM fermions. 
Even if the initial decay is  symmetric, e.g. $h\rightarrow VV$, if $V$ has comparable 
branchings into distinct final states then there will be events which contain different stable particles. Consider as an example
the $Z_4$ model of Eq.~(\ref{z4}) with two stable candidates $S$ and $\chi_{1}$. Provided the vector is much heavier than 
the DM candidates, the vector will decay via $V \rightarrow SP$ and $V\rightarrow \chi_{1}^\dag\chi_{1}$
with partial widths proportional to the square of the charges. Thus, the yield of $SP$ vs. 
$\chi_{1}^\dag\chi_{1}$ will be 4:1. The signature in this case would be  $\ell^+ \ell^- \not \!\!E$ with 
missing energy coming from two $S$ particles on one side of the event and two $\chi_{1}$ particles on the other. 
There are of course many other possibilities and in general it would be worthwhile to investigate the extent to
which the properties (mass, spin, etc.) of the different stable states can be determined. 
There has been recent work in this direction in Ref.~\cite{asymmetric}.
Along these lines, an interesting question is whether colliders have something to say regarding the
nature of the discrete symmetry at hand, i.e. can one distinguish between, say, $Z_2$ and $Z_3$ symmetries?
This question has received attention recently in Ref.~\cite{Z3Z2}, where the authors study the difference in the possible
decays of parent particles obeying a $Z_2$ vs. $Z_3$ symmetry and how such differences manifest in the invariant mass 
distributions of the outgoing SM particles. Another avenue of determining the underlying discrete gauge symmetry
in these models is to extract the charges of the Higgs boson and matter fields from collider data.
Measuring specific decay widths and branching ratios of e.g. the gauge boson into matter states can provide some 
information in this regard.
 
If the dark sector is very light compared to $m_h$, say at a scale of $O({\rm GeV})$ or less, 
the decay products will be highly boosted, leading to a novel ``lepton jet'' signature \cite{leptonjet}: 
multiple highly-collimated groups of leptons in the final state. In fact, it has recently been 
suggested that a SM Higgs decaying to lepton jets could have been missed at the LEP experiments \cite{hide}, allowing
for a relaxed bound on the SM Higgs mass. As in the unboosted case, various aspects of the 
collider phenomenology, such as event topology, lepton multiplicity, and missing energy, is highly dependent on
the spectrum in the dark sector. Many of the suggested search strategies in \cite{hide},
and past searches at LEP and Tevatron will be relevant 
for the class of models considered here. In particular, if we restrict to a Higgs mass above the LEP bound, $m_h > 114$ 
GeV \cite{lep}, searches at the Tevatron for `dark photons'~\cite{darkphoton}, NMSSM pseudoscalars~\cite{NMSSM}, 
trileptons~\cite{trilepton}, and same-sign dileptons~\cite{ssdl}
have the potential to constrain particular models and spectra. 

To summarize, if moderate couplings to the portals on the order of the bottom quark Yukawa exist, there is an 
opportunity to probe the dark sector at high energy colliders via the decays of the SM Higgs boson. 
Looking forward, it seems useful to proceed
in two directions due to the model dependence of the signatures. First, 
it would be worthwhile to study particular benchmark models containing some of the more complex signatures 
of multi-lepton, multi-jet final states with missing energy. Complementary to this, the parameter space of the more minimal 
models, particularly those with a single matter field, can be thoroughly explored.

{\bf Discussion.} Gauge symmetry provides a plausible origin for the 
discrete symmetries that can stabilize DM. In this paper we have surveyed a class of DM 
models based on an abelian $Z_N$ discrete gauge symmetry. We have investigated 
models with minimal field content, multi-field $Z_N$ models, and models
with chiral matter. We have found that $Z_N$ symmetries may lead to 
multiple stable DM candidates when $N$ is not prime. The dark sector may 
couple to the SM through the kinetic mixing and Higgs portals, and  we have 
given an outline of the basic phenomenology of this scenario, highlighting 
some novel direct detection and collider signatures. In particular, the SM 
Higgs boson may provide a portal through its decays to new states in the dark sector.

Our focus in this work has been on the gauge origin of the discrete symmetries 
and its potential impact on DM physics. 
There are a number of avenues for future work. 
Regarding indirect probes of these models, 
 DM annihilation in the galaxy and in the sun can lead to 
novel astrophysical signatures. As is well-known by now, a very light 
gauge boson provides a `dark force'~\cite{darkforce} 
that can potentially explain the cosmic ray signatures of PAMELA~\cite{pamela} and FGST~\cite{fermi}. 
In this context, there is scope for the exploration of the low energy phenomenology of these models
in fixed target~\cite{slacfixed,BPRfixed} and $e^+e^-$ experiments~\cite{Bfac}. 
Also, DM annihilation in the sun to light metastable
mediators, such as dark Higgs or gauge bosons, can yield electromagnetic signatures
in gamma-ray telescopes~\cite{gamma} such as FGST~\cite{fermi}. In terms of cosmology, as recently pointed out in~\cite{bbngev}
GeV-scale states from a dark sector can help to resolve the lithium problem of Big Bang Nucleosynthesis.  
It would be valuable to study these and other connections with cosmology and astrophysics for the DM models presented
here. 

In terms of model-building, we have not attempted to address the electroweak 
hierarchy problem, or the additional naturalness issues that come with scalars in the dark sector. 
In this regard, it would be interesting 
to adapt these models to a supersymmetric framework along the lines of \cite{leptonjet,abelianHS}
or a warped-extra-dimensional framework~\cite{warped}.  This can dramatically alter the collider 
phenomenology, as now one may use the supersymmetric or Kaluza-Klein states as connectors to the dark 
sector. 

We have focused on DM in a hidden sector, 
but one could also consider new U(1) gauge symmetries
under which the SM is charged. There are a few anomaly-free global U(1)
symmetries that one may gauge  within the SM: $B-L$ (with right-handed neutrinos) 
or a difference in lepton numbers, e.g. $L_e-L_\mu$~\cite{zp}. However, more general U(1) symmetries can 
be gauged provided new fermion representations are added to cancel anomalies, 
and these could be used to obtain discrete gauge symmetries.

Stability (at least on cosmological time scales) is one of the few known properties of DM and hints at a  
new symmetry in the dark sector. Should experimental evidence for DM be found, new stabilization 
symmetries may be required to understand the data. 
In the meantime, exploring new possibilities for these symmetries may suggest new 
phenomena associated with DM and therefore new search strategies. 
Our investigation here illustrates this basic observation, as
we have been led by symmetry considerations to predict 
new states, 
new interactions, and ultimately new signatures for direct detection and 
collider experiments.


{\bf Acknowledgments:} 
We thank Maxim Pospelov, Josef Pradler, and Adam Ritz for helpful discussions and 
comments on the manuscript. Research at the Perimeter Institute
is supported in part by the Government of Canada through NSERC and by the Province of Ontario through MEDT.


\end{document}